\documentstyle[12pt]{article}
\newlength{\pubnumber} 

\catcode`\@=11
\def\section{\@startsection{section}{1}{\z@}{3.5ex plus 1ex minus .2ex}
 {2.3ex plus .2ex}{\large\bf}}
\def\subsection{\@startsection{subsection}{2}{\z@}{2.3ex plus .2ex}
 {2.3ex plus .2ex}{\bf}}

\begin{document}

\begin{titlepage}
\samepage{
\setcounter{page}{1}
\rightline{McGill/93-34}
\rightline{\tt hep-th/9309126}
\rightline{August 1993}
\vfill
\begin{center}
 {\Large \bf Conformal Invariance\\
 and Degrees of Freedom\\
  in the QCD String\\}
\vfill
 {\large Keith R. Dienes\footnote{E-mail address:
                         dien@hep.physics.mcgill.ca}
   and Jean-Ren\'e Cudell\footnote{ E-mail address:
                         cudell@hep.physics.mcgill.ca}\\}
\vspace{.15in}
 {\it  Department of Physics, McGill University\\
3600 University St., Montr\'eal, Qu\'ebec~H3A-2T8~~Canada\\}
\end{center}
\vfill
\begin{abstract}
  {\rm We demonstrate that the Hagedorn-like growth of the number of observed
   meson states can be used to constrain the
   degrees of freedom of the underlying effective QCD string.
   We find that the temperature
   relevant for such string theories is not given by the usual
   Hagedorn value $T_H\approx 160$ MeV, but is considerably higher.
   This resolves an apparent conflict with the results from
   a static quark-potential analysis, and suggests that conformal
   invariance and modular invariance are indeed reflected in the
   hadronic spectrum.  We also find that the $D_\perp=2$ scalar
   string is in excellent agreement with data.}
\end{abstract}
\vfill}
\end{titlepage}

\setcounter{footnote}{0}

It is by now a well-established fact that many aspects of hadronic physics
can be successfully modelled by strings.  Indeed, an effective ``QCD string''
theory would simultaneously explain the existence of Regge trajectories,
linear confinement, the exponential rise in hadron-state densities,
and $s$- and $t$-channel duality.  In fact, it has recently been proposed
that string modular invariance might even explain
relative meson/baryon abundances \cite{FRCD}.

Over the years many different string models have been proposed for
describing the QCD color flux tube deemed responsible for quark confinement
in mesons:   early examples include the scalar (Nambu) string, the Ramond
string, and the Neveu-Schwarz (NS) string, and more recent examples
\cite{newstrings} include Polyakov's ``rigid string'', Green's
``Dirichlet string'', and the Polchinski-Strominger
effective string.    While
all of these models are endowed
with a certain number $D_\perp$ of bosonic degrees of freedom on the
two-dimensional string worldsheet (corresponding to the vibrational and
rotational string degrees of freedom in the external spacetime),
the early models place additional,
 {\it purely internal}\/ degrees of freedom on the string worldsheet.
The later string models instead introduce effective interactions for the
spacetime degrees of freedom which alter their short-distance behavior.

Conformal invariance nevertheless plays an important role in each case.
In the early models, conformal invariance is exact (though classical),
and indeed an infinite array of models of
this type can be constructed by placing on the string worldsheet virtually
any two-dimensional conformal field theory (CFT).  The quantum excitations of
the corresponding new worldsheet fields should then also produce new hadronic
states.  Conformal invariance in some of the more recent models, by contrast,
is an {\it effective}\/ symmetry, valid in the long-string limit.  The
effective CFT's which emerge in this limit are nevertheless equally
responsible for reproducing the salient features of hadronic spectroscopy.

A comparison with data is therefore necessary in order to constrain the class
of strings appropriate for modelling hadronic physics, or, more generally, to
test the validity of the string approach by determining the extent to which
conformal invariance is actually reflected in the observed hadron spectrum.
In particular, we shall use data to constrain the worldsheet central charge
$c$ of the effective QCD string, since this parameter is a unique measure of
the degrees of freedom in a general two-dimensional theory.  Most previous
efforts in this direction have employed the static quark potential method, and
favor values $c\approx 2$.
We propose using the well-known Hagedorn-like growth in the number
of meson states as an independent method constraining
not only the total central charge, but also its distribution between spacetime
and internal degrees of freedom.
While the value $T_H\approx 160$ MeV commonly taken for the
Hagedorn temperature \cite{Hagedorn} is too low to agree with the above
results (yielding $c\approx 7$), we will show that for comparisons
with strings an alternative treatment is necessary.
This will resolve the conflict between
these two methods, and in so doing provide experimental evidence
that the QCD string possesses conformal -- and indeed
modular -- invariance.  As a by-product, we will also find that
the $D_\perp=2$ scalar string is in excellent agreement with data.

Let us first review the basic ideas behind the static potential $V(R)$
between two quarks a distance $R$ apart.  Modelling confinement as a
color flux tube, one can show \cite{Luscheretal}
that this potential takes the exact form
\begin{eqnarray}
   V(R)~&=&~ \left[ (\sigma R)^2 + {M_0}^2 \right]^{1/2}~~~~~
    \nonumber\\
     ~&\sim&~ \sigma R + ({M_0}^2 /2\sigma)\,R^{-1}  + {\cal O}(R^{-2})~,
\label{Luscherform}
\end{eqnarray}
where $\sigma$ is the string tension of the flux tube
and ${M_0}^2\leq 0$ is a constant independent of $R$.
While the first (linear) term in the large-$R$ expansion of $V(R)$
represents the classical energy in the effective string,
the second ``pseudo-Coulomb'' term
is an attractive universal quantum correction (or Casimir energy)
which arises due to transverse zero-point vibrations of the string.
As such, this term is to be distinguished from the true
attractive Coulomb term which has the same form and which
arises at short distances from gluon exchange.
The form of the exact string result (\ref{Luscherform})
indicates that while $\sigma R$ plays the role of a string
``momentum,'' the quantity $M_0$ appears as the string ``rest mass''
(or ground-state energy) \cite{BN}.
The fact that ${M_0}^2\leq 0$ (or equivalently that the
long-distance pseudo-Coulomb term is attractive) implies that
the ground state of the effective QCD string is tachyonic, yet
this causes no inconsistency in the large-$R$ limit \cite{Olesen}.

If we assume the dynamics of the color flux tube to be modelled by a
two-dimensional conformal field theory, then this ground state energy $M_0$ and
the central charge $c$ of the corresponding worldsheet theory are related
by
\begin{equation}
     \alpha'\,{M_0}^2 ~=~ h - c/24~
\label{groundstate}
\end{equation}
where $\alpha'\equiv (2\pi\sigma)^{-1}$ is the Regge slope, and $h$ is the
conformal dimension of that primary field
in the worldsheet theory which produces the ground state.
In most cases this primary field is merely the identity field with
$h=0$, so that the coefficient of the
pseudo-Coulomb term directly yields the corresponding central
charge.  In all other cases, however, the coefficient of this term
yields information concerning only the {\it difference}
$h - c/24 \equiv - \tilde c/24$.
This is dramatically illustrated in the Ramond string:  here
$h = c/24 = (D-2)/16$, whereupon $\tilde c=0$, the ground state is massless,
and the long-range pseudo-Coulomb term is absent.

By fitting the parameters of
Eq.~(\ref{Luscherform}) to
heavy-quark spectroscopic data and/or results from lattice QCD, many
authors \cite{potentialfits} have attempted to determine the ground-state
energy ${M_0}^2$
and thereby the central charge of the QCD string.
While the string tension $\sigma$ is generally found to be in good agreement
with the Regge-trajectory value $\alpha'\approx 0.85$ (GeV)$^{-2}$,
values of $\tilde c$ have been obtained throughout
the range $0<\tilde c\leq 4$, clustering near $\tilde c\approx 2$.
Perhaps the largest source of error in these methods is the fact
that they rely upon a full separation of the effects of the long-range Coulomb
term from those of the true Coulomb interaction:  while the latter
are {\it a priori}\/ unrelated to $\tilde c$, fits to spectroscopic data
and/or lattice QCD results undoubtedly contain their contributions.
Furthermore, as discussed, $\tilde c$ is not always a true
measure of the degrees of freedom in the string worldsheet theory.

There exists, however, a different approach towards determining the central
charge $c$, one which avoids all of the above difficulties and is
complementary to that involving the static quark
potential.  In string theory (or more generally in any CFT), the number or
degeneracy of states $g_n$ at any excitation level $n$ is given by the
coefficients in a certain polynomial $\chi_h(x)$ called
the {\it character}\/ of the sector $[h]$ of the worldsheet CFT:
\begin{equation}
     \chi_h(x) ~= ~ x^{-\tilde c/24}\, \sum_{n=0}^\infty \,g_n \,x^n~.
\label{character}
\end{equation}
Since the spacetime mass $M_n$ of a given string excitation level $n$
is given by $\alpha'{M_n}^2 = n - \tilde c/24$, we see that the ground
state energy ${M_0}^2$ in each sector $[h]$ is indeed given by
Eq.~(\ref{groundstate}).  Thus, the static quark potential method,
by fitting $M_0$, is essentially a test of the string $n\to 0$ limit.
However, as has been well-known from the earliest days of the dual-resonance
models, information can also be extracted from the {\it high}\/-energy limit
$(n\to \infty)$, for in this limit string theories predict an exponential
rise in the degeneracy of states $g_n$ with excitation number $n$:
\begin{eqnarray}
      g_n~&\sim&~ A \,
      \lbrack C^2(n-\tilde c/24)\rbrack^{-B}\,
  e^{C\sqrt{n-\tilde c/24}}\nonumber\\
      ~&=&~ A \, \left(M/T_H\right)^{-2B}\,e^{M/T_H}~.
\label{exporise}
\end{eqnarray}
Here $A$, $B$, and $C$ are constants, with $T_H\equiv (C\sqrt{\alpha'})^{-1}$.
The form of these expressions demonstrates that $T_H$ is the famous Hagedorn
temperature \cite{Hagedorn}, a critical temperature beyond which the
partition functions of such theories (and indeed all of their thermodynamic
quantities) cannot be defined.  The exponent $B$ also has profound physical
consequences.  Since the internal energy of such a hadronic system near
$T_H$ grows as $U(T)\sim (T_H-T)^{2B-3}$
\cite{Frautschi}, we see that $T_H$ is a true maximum temperature if
$B\leq 3/2$, and merely the site of a
second-order QCD phase transition otherwise.

What makes these observations useful for our purposes, however,
is the fact that $T_H$ in Eq.~(\ref{exporise}) is directly related to the
total central charge of the underlying CFT:
\begin{equation}
       c ~=~ {3\over 2\pi^2} \,\left(\alpha'  \,{T_H}^2\right)^{-1}~.
\label{Hagcentral}
\end{equation}
Indeed, this result holds independently of the sector $[h]$
in which our string states are presumed to reside, yielding a
value for the true central charge $c$ rather than $\tilde c$
and providing
a test independent of the static quark potential method.
Furthermore, the exponent $B$ can also be interpreted in terms
of an underlying string theory, for $B$ is universally related to $D_\perp$,
the effective number of spacetime dimensions for transverse
string oscillations (or equivalently the number of
uncompactified bosonic degrees of freedom in the worldsheet CFT \cite{Dperp}):
\begin{equation}
          B~=~ {\textstyle{1\over 4}} \,(3 + D_\perp)~.
\label{Bdef}
\end{equation}
Thus, a fit of Eq.~(\ref{exporise}) to hadronic data
yields information concerning not only the
total number of degrees of freedom, but also their effective
distribution between spacetime and internal excitations.

In practice, however, a number of subtleties arise which
must be addressed before an adequate fit can be performed.
First, by tabulating the experimentally measured
masses $M_i$ and widths $\Gamma_i$ of observed
mesons \cite{PPDB}, we have
calculated the density $\rho_{\rm exp}(M)$ of meson states, where
$\rho_{\rm exp}(M) \equiv \sum_i W_i {\cal S}(M;M_i,\Gamma_i)$.
Here $W_i\equiv\gamma_i(2I_i+1)(2J_i+1)$ is the number of states per
resonance (with $\gamma_i=1$ for charge self-conjugate states and
$\gamma_i=2$ otherwise), and ${\cal S}$ represents a statistical
distribution function.  The result is plotted in Fig.~1, with
error bars determined by varying ${\cal S}$ between Breit-Wigner, Gaussian,
and fixed-width ($\Gamma_i=200$ MeV) distributions.  These errors also
include an estimate of the uncertainties resulting from the possible
subtraction of quark masses.  While this density clearly experiences
the predicted exponential growth over much of the plotted mass range, we
see that the rate of growth sharply diminishes beyond 1.7 GeV.  This is
attributable to experimental difficulties, for at higher energies
it becomes harder to distinguish mesons from background.  Indeed, a
comparison with explicit quark-model calculations \cite{Isgur} shows
1.7 GeV to be the first energy where fewer mesons are observed than predicted.
At the other extreme, the data below $0.3$ GeV depends on pion
contributions whose small masses result from
chiral symmetries which string theory is not expected to model.
Thus, we shall limit our attention to the experimental meson
data for $0.3\leq M\leq 1.7$ GeV.

This in turn requires a more sophisticated treatment of the
theoretical string predictions than was sketched above.  In particular,
taking $\alpha'\approx 0.85$ (GeV)$^{-2}$, we see that a mass near
$1.7$ GeV corresponds only to a string excitation of $n\approx 3$.
While the $n\to\infty$ asymptotic function quoted in Eq.~(\ref{exporise})
is remarkably accurate even for relatively small values of $n$,
it differs from the true string degeneracies $g_n$ for $n\leq 3$
by as much as 95\% for the scalar string.
We thus require from string theory a more precise functional form,
one which includes a sufficient number of subleading terms
so that the true values of $g_n$ for the known strings are accurately
reproduced.  It is a straightforward matter \cite{HR} to determine
the first set of these subleading terms, however, and together these yield
the following improved ``asymptotic'' form for the state degeneracies
$g_M$ and corresponding density $\rho(M)$:
\begin{eqnarray}
    g_M~&\sim&~\sqrt{2\pi}\,A~\xi^\nu\,I_{|\nu|}(\xi)
           ~~~~{\rm where}~ \xi\equiv M/T_H~ \nonumber\\
     &\equiv &~ (2\alpha'M)^{-1}\,\rho_{\rm string}(M)~.
\label{rhostring}
\end{eqnarray}
Here $I_{|\nu|}$ is the modified
Bessel function of order $|\nu|$, with $\nu\equiv 1/2-2B$;
note that this result reproduces Eq.~(\ref{exporise})
in the limit $M\to\infty$ \cite{density}.
By substituting the proper values of $A$, $B$, and $T_H$ for known strings,
we have verified that Eq.~(\ref{rhostring}) is indeed accurate over the
required range to within 2\%.  This is fortunate, since the forms of any
additional subleading terms are dependent on model-specific parameters
other than $B$ and $T_H$.

We then performed a fit
comparing the experimental and theoretical values
of $\int_{m_i}^{m} \rho(M) dM$ as a function of $m$, with $m_i=0.3$ GeV.
Our results are as follows.  Taking $B=5/4$ ({\it i.e.}, $D_\perp=2$)
resulted in a best fit with
\begin{equation}
      T_H ~=~ 300~{\rm MeV} ~~~\Longrightarrow~~~
      c~=~ 1.97~,
\label{fitresults}
\end{equation}
where we have taken $\alpha'=0.85$ (GeV)$^{-2}$.
This is remarkably close to the central charge $c=2$ of the $D_\perp=2$
scalar string, and demonstrates that a string picture is indeed consistent
with the data obtained from counting the numbers of hadronic states.
Note, in this regard, that the original 1967 fits by Hagedorn \cite{Hagedorn}
yielded $T_H\approx 160$ MeV, or $c\approx 7$,
in clear contradiction with the results from the static inter-quark potential.
However, the fundamental difference
is the functional form to which the fits are made,
for Hagedorn's bootstrap-motivated functional form \cite{Hagedorn}
$\rho(M) = a(M^2+\mu^2)^{-5/4} e^{M/T_H}$
with $\mu\approx 500$ MeV
has no connection to the string-theory result in Eq.~(\ref{rhostring})
and implicitly assumes $B=7/4$ in the $M\to\infty$ limit.
Indeed, it is only upon taking the string-motivated
functional form in Eq.~(\ref{rhostring}) that agreement is obtained.

In light of interesting recent proposals \cite{Caselle}
for {\it non}-scalar QCD strings with values $B\not= 5/4$, it is
nevertheless important to place general constraints in $(B,T_H)$ parameter
space.
In Fig.~2, the singly- and doubly-shaded regions indicate the
$(B,T_H)$ values for which
fits to the data with a $\chi^2/{\rm d.o.f.}\leq 1$ can be obtained:
the dashed line indicates the central value of the fits ({\it i.e.},
the ``best fits''), and the heavy solid lines indicate the border of the
region allowed by string theory (corresponding to the constraints
$B\geq 3/4$, $c\geq D_\perp$ \cite{Zamo}).  We see that there indeed
exists a region of overlap between the data-allowed and string-allowed
regions;  furthermore, the ``best fit'' line passes directly through this
overlap region in the range $3/4\leq B\leq 5/4$.
In the case of scalar strings for general $D_\perp$ (which
lie along the upper curved boundary of this string region), we
see that the ``best fit'' line intersects this scalar string line
almost exactly at the expected value $B=5/4$, or $D_\perp=2$.

The overlap region can be further narrowed if we constrain the
normalization constant $A$ in Eq.~(\ref{rhostring}).
Within the allowed region in Fig.~2, we find that
$A_{\rm fit}$ ranges from ${\cal O}(1)$ to $\geq {\cal O}(10^3)$.
String theory, however, places the precise limit
$A_{\rm string} \leq \sqrt{2\pi}\, (4\pi\alpha'{T_H}^2 )^\nu$,
with equality occurring for the scalar string.  Thus, if we assume that
there are $36$ independent strings which contribute to the
$(u,d,s)$-quark meson spectrum (corresponding to 36 quark degrees of
freedom:  9 possible quark/anti-quark flavor combinations, 4 spin states,
and one color singlet state), we can require
$r\equiv A_{\rm fit}/A_{\rm string}^{\rm max} \leq 36$.
This then restricts us to the lower (doubly-shaded) portion of
the data-allowed region in Fig.~2.
In Fig.~3 we have plotted versus $(c,D_\perp)$ that region of
allowed parameter space which satisfies all three of these string
constraints.  The best fit line is superimposed.

We have already seen in Eq.~(\ref{fitresults}) that the $c=D_\perp=2$
scalar string lies directly on the best fit line;  we now see from
Fig.~2 that this point is also exactly on the $r= 36$ border.
This implies that the scalar string should
accurately model the {\it absolute}\/ numbers of
states in the meson spectrum, and not merely their rate of growth.
In Fig.~1, for example, we have superimposed on
the meson data the actual numbers of states predicted by $36$ copies of
the scalar string (solid line). The agreement is excellent.

In conclusion, we have seen that the density of meson states
is consistent with string-theoretic predictions;  moreover,
estimations of the central charge of the QCD string obtained via measurements
of the appropriately-defined Hagedorn temperature are now consistent with those
independently obtained via static quark potential methods.  This latter
agreement is especially
significant, for these two methods depend separately on quantities which
are {\it a priori}\/ independent:  the rate of growth of the numbers
of mesons, and the energy of the ground state of the corresponding flux tube.
Indeed, as originally noticed in Ref.\ \cite{oldmodular},
only an underlying effective two-dimensional conformal invariance ---
in particular, modular invariance and the associated symmetry under
$x\to -1/x$ in Eq.~(\ref{character}) --- serve to relate them.  Our results
thus constitute strong additional evidence that the confinement phase
of QCD is consistent with an effective
string theory in which conformal symmetry and
modular invariance play a significant role.
This certainly warrants further study.

\bigskip
\medskip
\leftline{\large\bf Acknowledgments}
\medskip

We are pleased to thank A. Mironov for initial discussions, and M. Li and
R. Myers for comments on the manuscript.  This work was supported in part by
NSERC (Canada) and FCAR (Qu\'ebec).

\bigskip
\bigskip

\bigskip
\bibliographystyle{unsrt}

\begin{thebibliography}{99}
\bibitem[1]{FRCD}
  P.G.O. Freund and J.L. Rosner, {\it Phys.\ Rev.\ Lett.}\/ {\bf 68},
  765 (1992);  J.R. Cudell and K.R. Dienes, {\it Phys.\ Rev.\ Lett.}\/
  {\bf 69}, 1324 (1992).
\bibitem[2]{newstrings}
  A.M.\ Polyakov, {\it Nucl.\ Phys.} {\bf B268}, 406 (1986);
  M.B.\ Green, {\it Phys.\ Lett.}\/ {\bf B266}, 325 (1991);
  J. Polchinski and A. Strominger, {\it Phys.\ Rev.\ Lett.}\/
  {\bf 67}, 1681 (1991).
\bibitem[3]{Hagedorn}
  R. Hagedorn, {\it Nuovo Cimento} {\bf 56A}, 1027 (1968).
\bibitem[4]{Luscheretal}
  O. Alvarez, {\it Phys.\ Rev.} {\bf D24}, 440 (1981);
  M. L\"uscher, K. Symanzik, and P. Weisz,
  {\it Nucl.\ Phys.} {\bf B173}, 365 (1980).
\bibitem[5]{BN}
  L. Brink and H.B. Nielsen, {\it Phys.\ Lett.}\/ {\bf 45B}, 332 (1973).
\bibitem[6]{Olesen}
  P. Olesen, {\it Phys.\ Lett.}\/ {\bf 160B}, 144 (1985).
\bibitem[7]{potentialfits}
  See, {\it e.g.}, Ref.~\cite{Olesen} and references cited therein;  also
  A.A. Bykov, A.V. Leonidov, and A.D. Mironov,
  {\it Mod.\ Phys.\ Lett.}\/ {\bf A4}, 125 (1989).
\bibitem[8]{Frautschi}
  S. Frautschi, {\it Phys.\ Rev.} {\bf D3}, 2821 (1971);
  K. Huang and S. Weinberg, {\it Phys.\ Rev.\ Lett.} {\bf 25}, 895 (1970);
  N. Cabibbo and G. Parisi, {\it Phys.\ Lett.} {\bf 59B}, 67 (1975).
\bibitem[9]{Dperp}
  The general relation is $B=3/4-k/2$ where $k$ is the
  modular weight of the characters $\chi$ associated with
  the worldsheet CFT \cite{HR}.  Each transverse spacetime
  dimension (and the Liouville mode, if present) contributes $k= -1/2$.
  There are, however, no contributions from
  {\it compactified}\/ bosonic worldsheet fields (or
  equivalently from the worldsheet fermions of the R/NS strings).
\bibitem[10]{PPDB}
  Particle Data Group,  K. Hikasa {\it et al.},
  {\it Phys.\ Rev.}\/ {\bf D45}, S1 (1992).
\bibitem[11]{Isgur}
  S. Godfrey and N. Isgur, {\it Phys.\ Rev.} {\bf D32}, 189 (1985).
\bibitem[12]{HR}
  G.H. Hardy and S. Ramanujan, {\it Proc.\ Lon.\ Math.\ Soc.} {\bf 17}, 75
  (1918);  I. Kani and C. Vafa, {\it Commun.\ Math.\ Phys.} {\bf 130},
  529 (1990).  Since $I_{-\nu}(x) = I_{\nu}(x)$ if $\nu\in {\bf Z}$ or
  $x\gg 1$, we restrict ourselves to Bessel functions with positive order
  in Eq.~(\ref{rhostring});  these are better-behaved in the $x\sim
  {\cal O}(1)$ and unphysical ($\nu\not\in {\bf Z}$) regions.
\bibitem[13]{density}
  Since the states in a non-interacting string theory are infinitely narrow
  and populate discrete energy levels $M_n$, their density is
  $\rho(M) = g_M\sum_n\delta(M-M_n) =
  g_M(2\alpha'M)\sum_n\delta(\alpha'M^2+\tilde{c}/24-n)$.
  We replace the latter sum of $\delta$-functions with unity in order
  to provide a more realistic estimate of the generic broadening effects
  that interactions would have on the string spectrum.
  Furthermore, in the asymptotic limit $n\to\infty$, this yields
  $\rho(M) = g_M/\Delta M$ where $\Delta M$ is
  the mass difference between adjacent string levels.
\bibitem[14]{Caselle}
  M. Caselle, R. Fiore, F. Gliozzi, P. Provero, and S. Vinti,
  {\it Phys. Lett.}\/ {\bf B272}, 272 (1991).
\bibitem[15]{Zamo}
  It is possible, however, to violate the constraint $c\geq D_\perp$
  through renormalization group effects: see, {\it e.g.},
  A.B. Zamolodchikov, {\it JETP Lett.}\/ {\bf 43}, 730 (1986).
\bibitem[16]{oldmodular}
  L. Brink and H.B. Nielsen, {\it Phys.\ Lett.}\/ {\bf 43B}, 319 (1973);
  W. Nahm, {\it Nucl.\ Phys.} {\bf B114}, 174 (1976).
\end{thebibliography}

\vfill\eject

 ~
\vfill
\noindent {\bf Fig.\ 1}:
Number of meson states with masses $\leq M$ as function of $M$,
compared with scalar-string result.
\eject

 ~
\vfill
\noindent {\bf Fig.\ 2}:
Data- and string-allowed values of $(B,T_H)$, as discussed in text.
Best data fits (dashed line) and $D_\perp=2$ scalar string point (dot) also
shown.
\eject

 ~
\vfill
\noindent {\bf Fig.\ 3}:
Values of $(D_\perp,c)$ satisfying both data and string constraints,
with best fit line superimposed and scalar string shown (dot).
\eject
\end{document}